\documentclass[cameraready]{Interspeech}
\usepackage{multirow}
\usepackage{subcaption}
\usepackage{booktabs, xcolor, colortbl}
\definecolor{bestgreen}{RGB}{232,245,233}
\definecolor{fairgreen}{RGB}{46,125,50}
\definecolor{unfairred}{RGB}{183,28,28}
\definecolor{unfairblue}{RGB}{48,99,186}
\definecolor{frontendcolor}{RGB}{255, 192, 203}
\definecolor{backendcolor}{RGB}{255, 192, 203}
\definecolor{losscolor}{RGB}{255, 192, 203}
\definecolor{augmentcolor}{RGB}{200, 230, 255}
\definecolor{datacolor}{RGB}{255, 235, 180}
\definecolor{enginecolor}{RGB}{255, 255, 203}
\definecolor{evalcolor}{RGB}{144, 238, 144}
\definecolor{trainercolor}{RGB}{210, 180, 140}
\definecolor{logcolor}{RGB}{255, 160, 160}
\definecolor{configcolor}{RGB}{173, 216, 230}

\newcommand{\hlconfig}[1]{\colorbox{configcolor}{#1}}
\newcommand{\hldata}[1]{\colorbox{datacolor}{#1}}

\newcommand{\hlbackend}[1]{\colorbox{backendcolor}{#1}}

\newcommand{\hltrainer}[1]{\colorbox{trainercolor}{#1}}
\newcommand{\hllog}[1]{\colorbox{logcolor}{#1}}

\newcommand{\symfrontend}[1]{\texttt{#1}}
\newcommand{\symbackend}[1]{\texttt{#1}}
\newcommand{\symloss}[1]{\text{#1}}
\newcommand{\symtrndata}[1]{\text{#1}}
\newcommand{\symtestdata}[1]{\text{#1}}

\title{DeepFense: A Unified, Modular, and Extensible Framework for Robust Deepfake Audio Detection}

\author[affiliation={1}, correspondingauthor]{Yassine}{El Kheir}
\author[affiliation={1},equalcontribution]{Arnab}{Das}
\author[affiliation={2},equalcontribution]{Yixuan}{Xiao}
\author[affiliation={3},equalcontribution]{Xin}{Wang}

\author[affiliation={1}]{Feidi}{Kallel}

\author[affiliation={1}]{Enes Erdem}{Erdogan}

\author[affiliation={2}]{Ngoc Thang}{Vu}

\author[affiliation={1}]{Tim}{Polzehl}

\author[affiliation={4}]{Sebastian}{M\"oller}

\address{
$^1$ German Research Center for Artificial Intelligence (DFKI), Germany \\
$^2$ University of Stuttgart, Germany \\
$^3$ National Institute of Informatics, Japan \\
$^4$ Technical University of Berlin, Germany
}
\email{yassine.el\_kheir@dfki.de}
\usepackage[
backend=biber,
style=ieee,
maxbibnames=10,
maxcitenames=10,
doi=false,isbn=false,url=false,eprint=false, date=year,url=false
]{biblatex} 
\defbibheading{bibliography}[\refname]{}


\newcommand{\deepfense}{DeepFense}

\usepackage{comment}

\begin{document}

\maketitle
\begin{abstract}


Speech deepfake detection is a well-established research field with different models, datasets, and training strategies. However, the lack of standardized implementations and evaluation protocols limits reproducibility, benchmarking, and comparison across studies. In this work, we present DeepFense, a comprehensive, open-source PyTorch toolkit integrating the latest architectures, loss functions, and augmentation pipelines, alongside over 100 recipes. Using DeepFense, we conducted a large-scale evaluation of more than 400 models. Our findings reveal that while carefully curated training data improves cross-domain generalization, the choice of pre-trained front-end feature extractor dominates overall performance variance. Crucially, we show severe biases in high-performing models regarding audio quality, speaker gender, and language. DeepFense is expected to facilitate real-world deployment with the necessary tools to address equitable training data selection and front-end fine-tuning.

\end{abstract}

\begin{figure*}[!t]
    \centering
    \includegraphics[width=\linewidth, trim=0 220 0 0, clip]{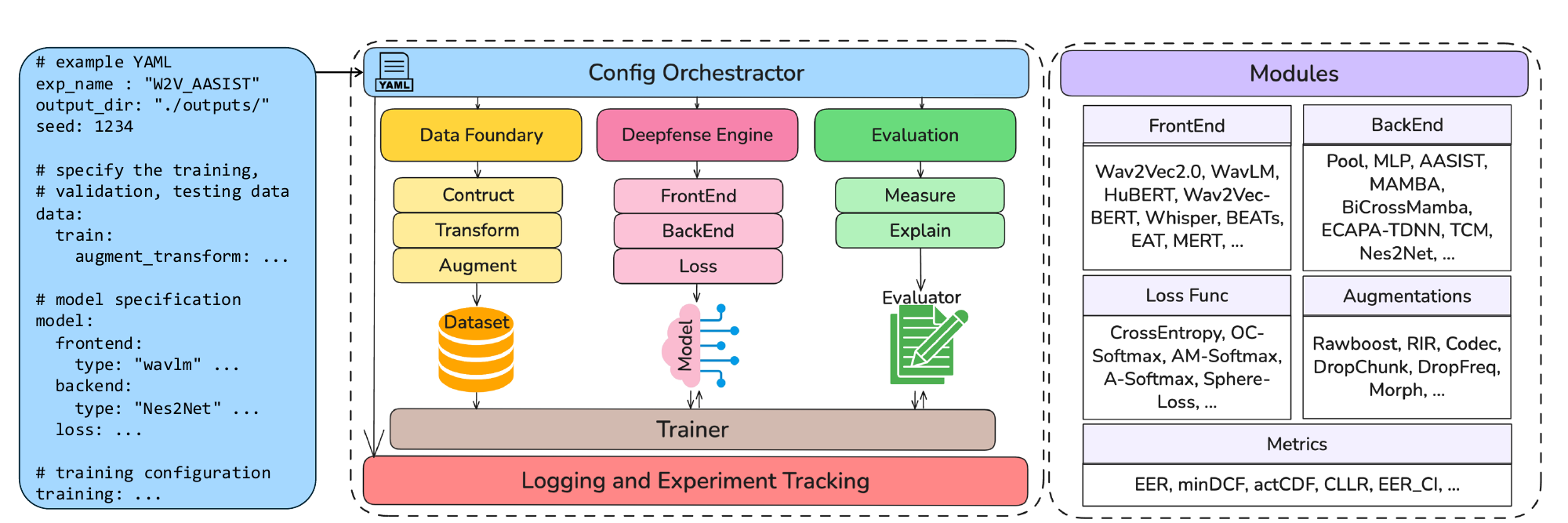}
    \caption{System architecture of the \deepfense{} framework, illustrating the configuration-driven data pipeline, modular model engine (front-end, back-end, loss), unified training loop, and evaluation components, with built-in logging and experiment tracking.}
    \label{fig:architecture}
\end{figure*}

\section{Introduction}
While speech synthesis technologies are essential in speech-based human-human and human-machine interaction, they are also being misused to forge speech deepfakes that threaten voice biometric systems~\cite{wu2017asvspoof} as well as human listeners~\cite{warren2024better}. The research community has increasingly devoted itself to the robust detection of speech deepfakes (and synthetic speech in general) over the last decade, creating numerous detector architectures~\cite{tak2020end,jung2022aasist,takAutomatic2022,elkheir25_interspeech}, databases~\cite{wangASVspoof2020,yiADD2023,liuASVspoof2023}, and training techniques~\cite{Tak2021, zhangOneClassLearningSynthetic2021}. Despite the numerous open-source projects that have implemented some of these research outcomes, many researchers, including ourselves, are seeking a full-fledged toolkit that covers many of the latest state-of-the-art models, provides recipes for diverse databases, and remains easy to use for implementing new research ideas and conducting large-scale experimental comparisons. 

In this paper, we present the \emph{\deepfense}\footnote{\href{https://deepfense.github.io/}{deepfense.github.io}} toolkit to answer this call. Targeting the research community, we are specifically motivated to improve the following aspects:
\begin{itemize}
    \item Fragmented implementations: Many building blocks, including feature extraction front-ends (e.g., Wav2Vec 2.0~\cite{wav2vec}), classification back-ends (e.g., AASIST~\cite{jung2022aasist}), and augmentations (e.g., RawBoost~\cite{Tak2021}), are implemented in different code repositories. We need extra time to glue these blocks together by adding wrappers around the models. Furthermore, we must manually download datasets distributed across various sources and re-invent the data IO tailored for each dataset.
    \item Hidden ingredients in recipes: When assembling blocks, we may inadvertently inherit recipes with varied ``hidden'' configurations: different padding and batching strategies, varied learning schedules, and diverse codebases (e.g., fairseq~\cite{ott_fairseq_2019} vs. Hugging Face~\cite{wolf_transformers_2020})\footnote{For example, Hugging Face uses a separate \texttt{FeatureExtractor} to normalize the waveform before feeding it to Wav2Vec2, which may be missed by users who use Fairseq (\url{https://github.com/huggingface/transformers/issues/28726}).}. Some leaderboards, such as Speech-DF-arena~\cite{11345101}, allow anyone to submit a system and evaluate it on a selected set of test sets. However, not all entries are open-sourced.
    \item Programming barrier: While there is a recently released toolkit called \emph{WeDefense}~\cite{zhang2026wedefensetoolkitdefendfake} dedicated to solving the two points above\footnote{DeepFense is developed independently and in parallel with WeDefense, without prior coordination or influence.}, it mixes Python and numerous Bash scripts, which complicates the debugging and extension procedures, with limited recipes and available models. While a pure-Python toolkit such as SpeechBrain~\cite{speechbrain} can mitigate this issue, its general-purpose design introduces abstractions and functions that may unexpectedly increase the cost of learning and customization\footnote{A good example is the AntiDeepfake project~\cite{antideepfake_2025}, which had to tailor the SpeechBrain core method \texttt{fit()} to add intra-epoch validation.}. 
\end{itemize}

The first two issues make it difficult to isolate the contribution of algorithmic innovations from implementation artifacts. \deepfense{} aims to address this by consolidating state-of-the-art architectures, recipes for major deepfake datasets, and a unified training-evaluation pipeline in a centralized repository. This differentiates \deepfense{} from tools that focus on either a single group of models (e.g., post-trained models in AntiDeepfake~\cite{antideepfake_2025}) or challenge baselines with recipes on specific datasets~\cite{liuASVspoof2023,yiADD2023}. With a pure Python/PyTorch framework, \deepfense{} is expected to be more friendly to the research community.

In short, the key contributions of \deepfense{} are as follows:
\begin{itemize}
    \item It is a pure-Python/PyTorch toolkit dedicated to deepfake detection. With an Apache 2.0 license, the toolkit is suitable for a wide range of users. 
    \item It provides \emph{more than 100 recipes and 400 pre-trained models} for different datasets and model architectures, which is the largest scale currently available.
    \item It facilitates our large-scale experimental comparison (\S~\ref{sec:exp}) that covers around \emph{100 systems, six training sets, and 13 test sets}, including speech and non-speech datasets.
\end{itemize}
Through this large-scale experiment, we showcase how \deepfense{} can be utilized to accelerate research progress by quickly conducting cross-dataset studies and building baselines.

In the rest of the paper, we describe the high-level design and functions provided by \deepfense{} (\S~\ref{sec:framework}). Through the experiment (\S~\ref{sec:replica}, \S~\ref{sec:exp}), we highlight our findings on domain overfitting, which is observed on the most well-known training sets in the field. We also discuss the impacts of front- and back-ends on detection performance. Based on the findings, we discuss future work and draw our conclusion (\S~\ref{sec:conclusion}).

\section{DeepFense Framework}
\label{sec:framework}
\subsection{Design principle}


As Figure~\ref{fig:architecture} illustrated,  \deepfense{} implements a modular architecture that separates experimental specification, data processing, model training, execution, and evaluation. The complete training-evaluation setup is specified in a \emph{single} human-readable file. These loosely coupled and highly modular components make the toolkit reusable while simplifying experiment execution and sharing among researchers. 

\begin{itemize}
    \item \hlconfig{Configuration Orchestrator}: It coordinates all framework components through YAML specifications, hence referred to as ``orchestrator''. 
    A single YAML file specifies the complete experiment: dataset paths, data preprocessing, model architecture (front-end, back-end, loss), training parameters, and evaluation protocols. An example is shown to the left of Figure.~\ref{fig:architecture}.
    The orchestrator parses YAML specifications and instantiates components via a registry system (\S~\ref{sec:design:register}). 

\item \hldata{Data Foundry}: it manages the data pipeline via four stages:
\begin{itemize}
\item \noindent\textbf{Construct:} Loads dataset metadata from Parquet files containing audio paths, labels, and optional metadata. This format enables efficient operations on large datasets.
\item \noindent\textbf{Transform:} Applies base preprocessing such as padding or cropping, resampling, and optional normalization. 
\item \noindent\textbf{Augment:} Applies stochastic transformations to the data during training. DeepFense supports both sequential augmentation (transformations applied in order) and parallel augmentation (randomly selecting one from a set). Each augmentation has a configurable probability.

\item \noindent\textbf{Dataset:} Constructs PyTorch DataLoaders with configured batch size, shuffling, and parallel loading. 
The Data Foundry outputs batched tensors of shape [batch, samples] that flow into the DeepFense Engine.
\end{itemize}

\item \hlbackend{DeepFense Engine}:
It implements the detection model as a composition of three components: \texttt{Front-end $\rightarrow$ Back-end $\rightarrow$ Loss}. The components available are listed in \S~\ref{sec:design:components}.

\item \hltrainer{Trainer}: 
The training loop iterates: (1) load batch from Data Foundry, (2) forward pass via DeepFense Engine, (3) backward pass and optimization, (4) log metrics, (5) evaluate on validation set periodically, (6) save checkpoints.

\item {\hllog{Logging and experiment tracking}}: DeepFense integrates with Weights \& Biases and TensorBoard for experiment tracking. All experiments generate structured output directories that contain model checkpoints, configuration files, training logs, evaluation results, and optional visualizations. 
This organization makes it easier to manage the experiments. 
\end{itemize}

\subsection{Supported models components \& losses}
\label{sec:design:components}

\deepfense{} integrates many SSL front-ends \symfrontend{Wav2Vec~2.0}~\cite{wav2vec}, \symfrontend{WavLM}~\cite{chenWavLM2022}, \symfrontend{HuBERT}~\cite{hsu2021hubert}, \symfrontend{EAT}~\cite{chen2024eat}, \symfrontend{MERT}~\cite{li2024mert}, all loaded via Hugging Face\footnote{\url{https://huggingface.co/}}, Fairseq\footnote{\url{https://github.com/facebookresearch/fairseq}}, and Unilm\footnote{\url{https://github.com/microsoft/unilm}} and other Speech/Audio models such as \symfrontend{Whisper}~\cite{radford2023robust}, \symfrontend{BEATs}~\cite{chen2022beats}, \symfrontend{Wav2Vec2-BERT}~\cite{chung2021w2v}. Front-ends can be used frozen, fine-tuned, or with learnable weighted-sum or attentive aggregation across layers.
Seven back-ends are supported (\symbackend{AASIST}~\cite{jung2022aasist}, \symbackend{ECAPA-TDNN}~\cite{desplanques20_interspeech}, \symbackend{RawNet2}~\cite{tak2020end}, \symbackend{Nes2Net}~\cite{liuNes2Net2025}, \symbackend{TCM}~\cite{truongTemporalChannel2024}, \symbackend{Multi-layer perceptron (MLP)\cite{el2025comprehensive}}, \symbackend{Pool} (Pooling), \symbackend{BiCrossMamba-ST}\cite{elkheir25_interspeech}), alongside four loss functions (\symloss{Cross-Entropy}, \symloss{OC-Softmax}~\cite{zhangOneClassLearningSynthetic2021}, \symloss{AM-Softmax}~\cite{wang2018additive}, \symloss{A-Softmax}~\cite{liuSphereFace2017}) and ten augmentation strategies including RawBoost~\cite{Tak2021}, room impulse response, and codec simulation.

\subsection{Modular extensibility}
\label{sec:design:register}
DeepFense uses a registry-based plugin architecture. Adding a new component (front-end, back-end, loss, augmentation) requires implementing a Python class inherited from the base interface and registering it with a decorator, as follows:

\begin{verbatim}
@BACKEND_REGISTRY.register()
class CustomBackend(BaseBackend):
    def forward(self, x):
        return embeddings
\end{verbatim}
Once registered, the component is available through configuration without modifying the core code. 

This configuration-driven approach provides several advantages for reproducibility and accessibility. Complete experiments are specified in a single human-readable file; hyperparameter searches involve only configuration changes without code modification; published results can be exactly reproduced from the configuration file; and systematic comparisons across methods become straightforward. 

\subsection{Available recipes \& pre-trained models}

Research in speech deepfake detection relies on several benchmark datasets that have evolved from controlled laboratory environments to ``in-the-wild'' scenarios. The ASVspoof series remains popular, with ASVspoof 2019 and 2021 (Logical Access and Deepfake tracks) providing a mix of synthetic speech and telephony-distorted speech utterances~\cite{liuASVspoof2023}. The Audio Deepfake Detection (ADD) challenges (2022, 2023) introduced more complex tasks, such as detecting partially manipulated segments and low-quality fakes~\cite{yiADD2023}. For broader generalization, researchers utilize the In-the-Wild dataset~\cite{muller24b_interspeech}, which consists of real-world clips from celebrities and politicians. Most recently, large-scale multilingual sets like MLAAD~\cite{mullerMLAAD2024} have expanded the task to over 50 languages, ensuring that models are robust against diverse linguistic and generative variations. A more recent dataset called CodeFake~\cite{wu_codecfake_2024} covers various diffusion and LLM-based synthetic speech. Additionally, we have added datasets used in the research community, such as HABLA \cite{florez2023habla}, \symtrndata{PartialSpoof} \cite{zhangPartialSpoofDatabaseCountermeasures2022}, \symtestdata{ODSS} \cite{yaroshchuk2023open}, \symtestdata{ReplayDF} \cite{muller2025replay})

Given the plethora of dataset choices, \deepfense{} ships \textbf{152 unique YAML recipes} and \textbf{456 trained checkpoints} (3 seeds each) publicly released on HuggingFace\footnote{\url{anonymous}}. This covers all front-end $\times$ back-end $\times$ training-dataset combinations across speech and non-speech tasks, and to our knowledge is the largest such collection for audio deepfake detection. Dataset Parquet files and download scripts for all supported benchmarks are also provided for one-command reproducibility.

\section{Replicating SOTA Results}
\label{sec:replica}
\begin{table}[!t]
\centering
\setlength{\tabcolsep}{3pt}
\caption{EER~(\%) for \textbf{Wav2Vec2} systems trained on \textbf{ASV19},
evaluated on out-of-domain test sets before and after applying DeepFense.
Best result per back-end in \textbf{bold}.}
\vspace{-2mm}
\label{tab:deepfense_asv19}
\resizebox{\columnwidth}{!}{
\begin{tabular}{l cc cc cc cc}
\toprule
& \multicolumn{2}{c}{ITW}
& \multicolumn{2}{c}{ASV5}
& \multicolumn{2}{c}{CodecFake}
& \multicolumn{2}{c}{Average} \\
\cmidrule(lr){2-3}\cmidrule(lr){4-5}\cmidrule(lr){6-7}\cmidrule(l){8-9}
Back-End & Orig. & Ours & Orig. & Ours & Orig. & Ours & Orig. & Ours \\
\midrule
AASIST   & 10.46 & \textbf{7.18}  & 19.24 & \textbf{18.94} & 38.79 & \textbf{34.36} & 22.83 & \textbf{20.16} \\
MLP      &  8.13 & \textbf{7.57}  & 18.76 & \textbf{15.43} & 35.76 & \textbf{34.44} & 20.88 & \textbf{19.15} \\
Nes2Net  &  7.06 &          7.73  & 22.06 &          24.19 & 39.34 & \textbf{36.21} & 22.82 & \textbf{22.71} \\
TCM      &  7.79 & \textbf{6.98}  & 18.85 &          19.40 & 36.00 & \textbf{34.77} & 20.88 & \textbf{20.39} \\
\bottomrule
\end{tabular}}
\end{table}

\begin{table}[!t]
\centering
\setlength{\tabcolsep}{7pt}
\caption{EER~(\%) for \textbf{Wav2Vec2} systems trained on \textbf{ASV5},
evaluated on the ASV5 test set.}
\vspace{-2mm}
\label{tab:deepfense_asv5}
\scalebox{1}{
\begin{tabular}{l cc}
\toprule
Back-End & Orig. & Ours \\
\midrule
AASIST           & 7.12 & \textbf{6.64} \\
Nes2Net          & 6.13 & \textbf{5.53} \\
BiCrossMamba-ST  & 6.01 &          6.34 \\
\bottomrule
\end{tabular}}
\end{table}

\begin{table}[htbp]
\centering
\setlength{\tabcolsep}{7pt}
\caption{EER~(\%) for \textbf{EAT} trained on EnvSDD and evaluated on \textbf{CodecFake-A3} test set.}
\label{tab:deepfense_envsdd}
\vspace{-2mm}
\scalebox{1}{
\begin{tabular}{l cc}
\toprule
Back-End       & Orig. & Ours \\
\midrule
Nes2Net        & 0.49 &          0.55 \\
AASIST         & 1.03 & \textbf{0.77} \\
MLP            & 0.95 & \textbf{0.91} \\
TCM            & 0.34 &          0.40 \\
BiCrossMamba-ST   & 0.75 & \textbf{0.44} \\
\bottomrule
\vspace{-10pt}
\end{tabular}}
\end{table}

\begin{figure*}[!t]
  \centering
  \includegraphics[width=\textwidth]{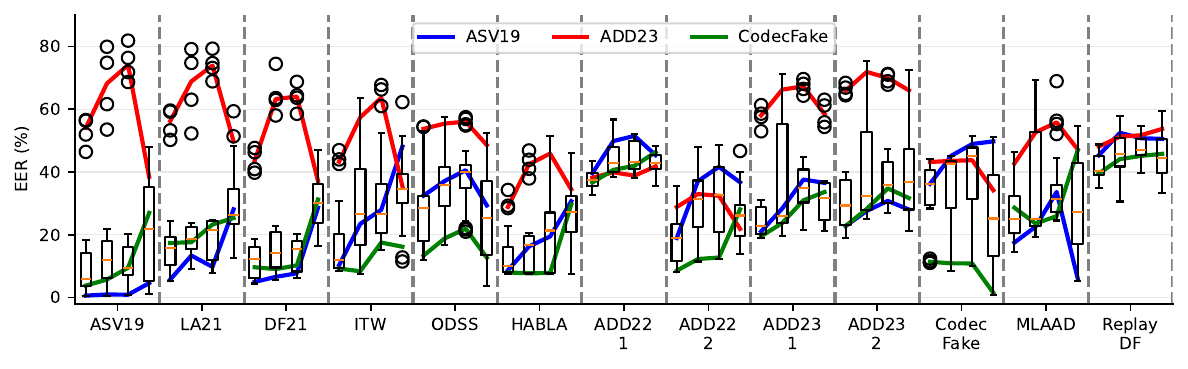}
  \vspace{-8mm}
  \caption{Boxplots of detection systems based on their EERs on the test set (x-axis). Each box groups 16 systems using the same front-end (from left to right: Wav2Vec2, WavLM, HuBERT, and EAT)  but varied training data and back-ends. The three lines correspond to mean EERs of systems using either the ASV19 (blue), ADD23 (red), or CodecFake (green) training sets.}
  \label{fig:overall}
\end{figure*}

\begin{figure*}[htbp]
  \centering
  \includegraphics[width=\linewidth]{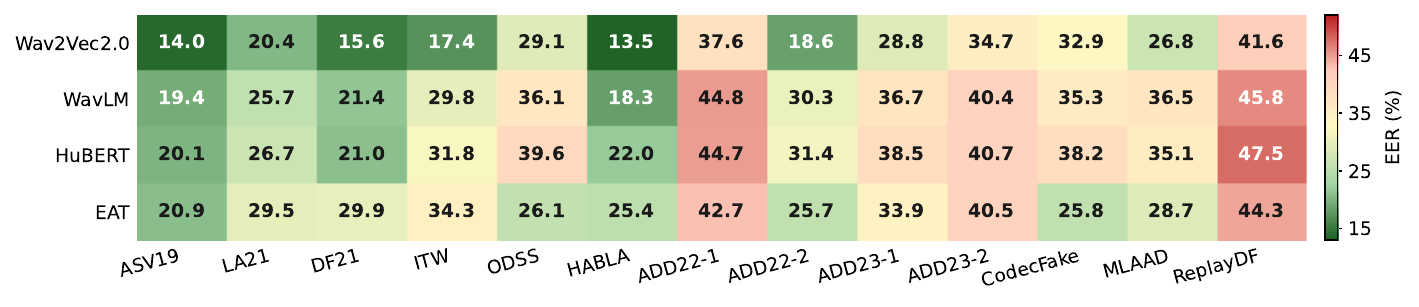}
  \vspace{-8mm}
  \caption{Mean EER~(\%) per front-end and evaluation benchmark, averaged over all back-ends and training sets.}
  \label{fig:heatmap}
\end{figure*}

To validate the correctness and reliability of the \deepfense{} framework, we assess its ability to replicate SOTA results from the literature. Specifically, we re-implement and evaluate several published systems under controlled conditions using \deepfense{}'s unified pipeline, ensuring identical preprocessing, optimizer configurations, and evaluation protocols.

Table~\ref{tab:deepfense_asv19} reports the EER (\%) for \symfrontend{Wav2Vec2}-based systems trained on ASVspoof 2019 (ASV19) and evaluated on three out-of-domain test sets: In-the-Wild (ITW), ASVspoof 5 (ASV5), and CodecFake. Across all four back-end classifiers (\symfrontend{AASIST}, \symfrontend{MLP}, \symfrontend{Nes2Net}, \symfrontend{TCM}), \deepfense{} consistently matches or outperforms the originally reported results, whether from the original paper, replicated, or from Speech-DF-Arena.


For instance, AASIST achieves an average EER reduction from 22.83\% to 20.16\%, and MLP improves from 20.88\% to 19.15\%, demonstrating that the framework does not introduce any degradation stemming from re-implementation. Table~\ref{tab:deepfense_asv5} further confirms these findings for systems trained in ASV5, where \deepfense{} reproduces competitive results for AASIST (7.12\% $\rightarrow$ 6.64\%) and Nes2Net (6.13\% $\rightarrow$ 5.53\%), achieving on-par or better performance compared to the baselines originally published.

Beyond speech-domain tasks, Table~\ref{tab:deepfense_envsdd} validates \deepfense{} on the non-speech CodecFake-A3 \cite{wu_codecfake_2024} test set using \symfrontend{EAT} front-end models trained on EnvSDD \cite{yin2025esdd}. The reproduced results closely match the original numbers across all back-ends, with \symfrontend{BiCrossMamba-ST} \cite{elkheir25_interspeech} achieving an improved EER of 0.44\% compared to the originally reported 0.75\%. 
In short, the results demonstrate that \deepfense{} can achieve on-par or better performance across diverse tasks, datasets, and architectures. 

\section{Large-scale Comparison}
Prior studies typically evaluate a single model or a small set of architectures on one or two datasets, making it difficult to disentangle the contributions of the front-end, back-end, and training data from implementation-specific choices~\cite{takAutomatic2022, jung2022aasist}.

A key motivation behind \deepfense{} is to enable systematic, large-scale comparisons that are otherwise infeasible without a unified framework. \deepfense{} makes this feasible by providing a single controlled pipeline where all configurations share identical preprocessing, optimizer settings, and evaluation protocols, ensuring that observed differences are attributable to modeling choices rather than experimental artifacts. Here, we utilize \deepfense{} to conduct a systematic, large-scale comparison exploring how the front-end, back-end, and training data affect the detection performance.

\label{sec:exp}
\vspace{-5pt}

\subsection{Setup}
We evaluate $4\ \text{front-ends} \times 4\ \text{back-ends} \times 6\ \text{training sets} = 96$ systems, each trained with seeds \{2, 42, 240\} (EER averaged over seeds).
The front-ends include \symfrontend{EAT}, \symfrontend{HuBERT}, \symfrontend{Wav2Vec2}, and \symfrontend{WavLM}, and the back-ends include \symbackend{AASIST}, \symbackend{MLP}, \symbackend{Nes2Net}, and \symbackend{TCM}. The training sets are \symtrndata{ASV19}, \symtrndata{ASV5}, \symtrndata{ADD23} (Chinese), \symtrndata{CodecFake}, \symtrndata{HABLA} (Spanish), \symtrndata{PartialSpoof}.
All systems are trained with cross-entropy loss, the Adam optimiser (lr=$10^{-6}$), inputs padded or cropped to 4\,s at 16\,kHz, early stopping on validation loss, and repeated 3 times with different seeds. The average across 3 seed experiments is reported per system. 

Our evaluation covers 13 test sets across four linguistic groups: English (\symtestdata{ASV19}, \symtestdata{ASVspoof21 LA} \symtestdata{(LA21)}, \symtestdata{ASVspoof21 DF} \symtestdata{DF21} \cite{liuASVspoof2023}, \symtestdata{ITW}, \symtestdata{CodecFake}, \symtestdata{ReplayDF}), Multilingual (\symtestdata{MLAAD}, \symtestdata{ODSS}), Chinese (\symtestdata{ADD22-1/2}, \symtestdata{ADD23-1/2}), and Spanish (\symtestdata{HABLA}). The motivation is to provide diverse and cross-lingual assessments. Note that we also include test sets that are not covered by the Speech-df-arena.

\subsection{Impact of front-end}

The results reveal that the choice of front-end is the most influential factor across all experimental dimensions. As shown in Figure.~\ref{fig:heatmap} and Table~\ref{tab:fe_be_avg}, \symfrontend{Wav2Vec2} achieves the lowest macro-average EER of $25.5\%$, outperforming the second best front-end (\symfrontend{EAT}, $31.4\%$), and is the best-performing front-end on 11 out of 13 evaluation datasets.
\symfrontend{HuBERT} consistently ranks last with a macro-average EER of $33.6\%$, suggesting that despite its strong performance on other tasks, its representations are less discriminative for deepfake artifacts under our evaluation conditions.
\symfrontend{EAT} and \symfrontend{WavLM} fall in the middle range ($31.4\%$ and $32.4\%$, respectively), with \symfrontend{EAT} notably outperforming \symfrontend{Wav2Vec2} on the ODSS and CodecFake benchmarks, suggesting some complementarity in the feature spaces these models capture.


The advantage of \symfrontend{Wav2Vec2} is most pronounced on the \symtestdata{ITW} dataset, where it achieves an EER of $17.4\%$ compared to $34.3\%$ for \symfrontend{EAT} and $31.8\%$ for \symfrontend{HuBERT}. The gap between \symfrontend{Wav2Vec2} and \symfrontend{EAT} is $16.9\%$, the largest observed across any single evaluation dataset.
This indicates that representations learned by \symfrontend{Wav2Vec2} are particularly robust to the diverse acoustic conditions present in real-world speech recordings.
In contrast, on the Chinese ADD benchmarks, the absolute EER values for all front-ends are considerably higher (ranging from $37.6\%$ to $44.8\%$ on ADD22-1), and the relative differences between front-ends are compressed. 
ReplayDF is universally the most difficult evaluation set, with EERs between $41.6\%$ (\symfrontend{Wav2Vec2}) and $47.5\%$ (\symbackend{HuBERT}), reflecting the more challenging scenario of replay effects on top of synthesis-based deepfake attacks.


\begin{table}[t]
\centering
\caption{Mean EER (\%) for each front-end (averaged over all back-ends \& training sets)
         and each back-end (averaged over all front-ends \& training sets). $\downarrow$}
\label{tab:fe_be_avg}
\vspace{-2mm}
\setlength{\tabcolsep}{5pt}
{
\begin{tabular}{lc|lc}
\toprule
Front-End & Mean EER (\%) & Back-End & Mean EER (\%) \\
\midrule
\textbf{Wav2Vec2} & \textbf{25.5} & \textbf{AASIST} & \textbf{30.3} \\
EAT               & 31.4          & Nes2Net         & 30.7          \\
WavLM             & 32.4          & TCM             & 30.8          \\
Hubert            & 33.6          & MLP             & 31.1          \\
\bottomrule
\end{tabular}}
\end{table}

\begin{figure*}[!]
  \centering
  \begin{subfigure}[t]{0.49\textwidth}
    \centering
    \includegraphics[width=\linewidth]{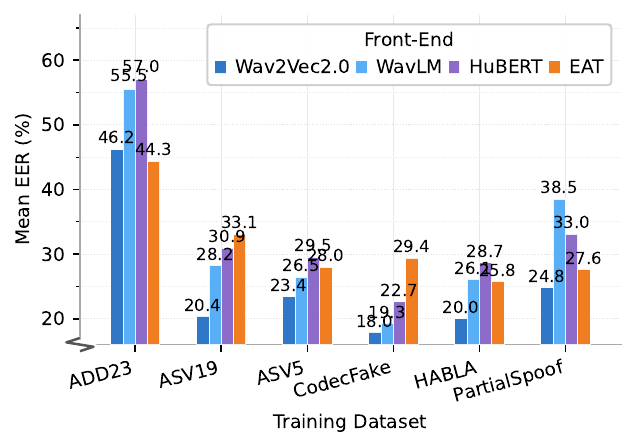}
    \vspace{-5mm}
    \caption{Mean EER per training dataset and front-end, averaged over all back-ends and 13 evaluation benchmarks.}
    \label{fig:train_effect}
  \end{subfigure}
  \hfill
  \begin{subfigure}[t]{0.49\textwidth}
    \centering
    \includegraphics[width=\linewidth]{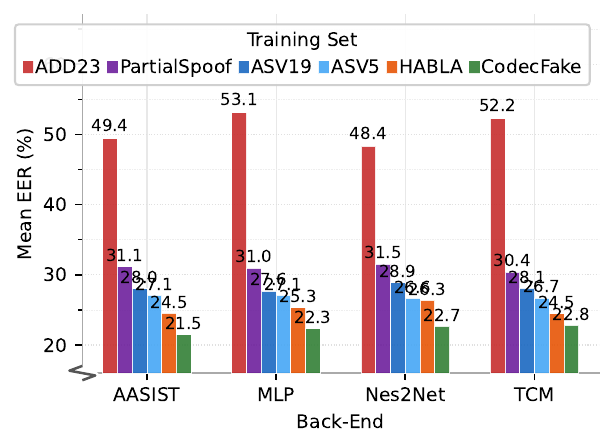}
    \vspace{-5mm}
    \caption{Mean EER for each back-end and training dataset combination, averaged over all front-ends and 13 evaluation benchmarks.}
    \label{fig:be_train}
  \end{subfigure}
  \vspace{-3mm}
  \caption{Mean EER analysis across training datasets, front-ends, and back-ends. (a) Front-end perspective. (b) Back-end perspective.}
  \label{fig:train_backend_combined}
\end{figure*}

\subsection{Impact of back-end}

In contrast to the front-end, the choice of back-end has a comparatively minor effect on detection performance. As shown in Tab.~\ref{tab:fe_be_avg}, aggregated across all front-ends and training datasets, the macro-average EER ranges from $30.3\%$ (\symbackend{AASIST}) to $31.1\%$ (\symbackend{MLP}), a difference of only $0.8\%$.
This result aligns with recent observations in the literature suggesting that the quality of the input representation dominates over the classification head architecture when strong SSL features are available~\cite{wangInvestigatingSelfSupervisedFront2022}.

Although \symbackend{AASIST} and \symbackend{Nes2Net} exhibit slight advantages, their improvements are not consistently maintained across different front-end configurations and training setups. Overall, the observed performance differences between back-end architectures remain relatively small compared to the variability introduced by front-end selection and training conditions, indicating that back-end effectiveness is strongly dependent on the overall system design.


\subsection{Impact of training dataset}

The training dataset has a dramatic effect on generalization. Table~\ref{tab:training} summarizes the macro-average EER per training dataset, aggregated over all front-ends and back-ends.

\begin{table}[!t]
\caption{Macro-average EER across all 13 evaluation datasets, per training set (mean $\pm$ std over all front-end $\times$ back-end).}
\vspace{-2mm}
\label{tab:training}
\centering
\small
\begin{tabular}{lcc}
\toprule
{Training Data} & {Mean EER} & {Std} \\
\midrule
CodecFake    & \textbf{22.3} & 0.047 \\
HABLA        & 25.2 & 0.034 \\
ASV5         & 26.9 & 0.029 \\
ASV19        & 28.2 & 0.052 \\
PartialSpoof & 31.0 & 0.056 \\
ADD23        & 50.8 & 0.063 \\
\bottomrule
\vspace{-10pt}
\end{tabular}
\end{table}

\noindent\textbf{CodecFake as the most transferable training set.}
Training on \symtrndata{CodecFake} yields the best macro-average EER ($22.3\%$) and the lowest variance ($\pm 4.7\%$) across all systems.
This is particularly striking because \symtrndata{CodecFake} achieves strong generalization not only on its in-domain test set, where it dramatically outperforms \symtrndata{ASV19}-trained systems ($8.6\%$ vs.\ $44.9\%$ EER on \symtestdata{CodecFake} eval), but also on entirely out-of-domain benchmarks including \symtestdata{ITW} ($12.8\%$, the best of any training set) and ODSS ($16.8\%$).
We hypothesize that codec-based artifacts introduce a broad and pervasive distortion signature that generalizes across synthesis pipelines, making it a powerful training condition even for non-codec deepfake systems.

\noindent\textbf{HABLA: surprising cross-lingual transfer.}
Despite being recorded entirely in Spanish, \symtrndata{HABLA}-trained models achieve the second-lowest macro-average EER ($25.2\%$) and show strong generalization to the English and multilingual evaluation sets ($24.6\%$).
The effective cross-lingual transfer suggests that the deepfake artifacts captured in \symtrndata{HABLA} are not language-specific but reflect more fundamental synthesis artifacts shared across languages.

\noindent\textbf{ADD23: a cautionary finding.}
Models trained on the Chinese \symtrndata{ADD23} dataset catastrophically fail to generalize, yielding a macro-average EER of $50.8\%$ equivalent to random guess.
Critically, \symtrndata{ADD23}-trained models also fail on the other Chinese evaluation sets (\symtestdata{ADD22} and \symtestdata{ADD23} eval, average EER $50.6\%$), which is worse than even the CodecFake-trained systems ($28.4\%$) on the same Chinese benchmarks.
This implies that the poor generalization is not due to a language mismatch, but rather to a mismatch in \emph{synthesis methods}.
This finding is a strong warning against assuming that in-language training data produces better cross-dataset generalization. 

\noindent\textbf{No single training set dominates all evaluation conditions.}
Consistent with prior cross-corpus studies~\cite{williams2008exploiting}, no training dataset achieves the best EER on all 13 evaluation sets simultaneously.
\symtrndata{ASV19} training, for example, achieves the lowest EER on the \symtestdata{ASV19} in-domain test ($1.8\%$) and performs competitively on \symtestdata{MLAAD} ($20.0\%$) and \symtestdata{ADD23-2} ($28.0\%$), but yields the worst performance on \symtestdata{CodecFake} ($44.9\%$) and \symtestdata{ADD22-2} ($33.5\%$).
\symtrndata{PartialSpoof} training achieves the lowest EER on \symtestdata{LA21} ($14.2\%$ averaged over front-ends), but is mediocre on most other benchmarks.

Furthermore, Figure.~\ref{fig:overall} illustrates that the performance on a test set is the best when the training set matches the domain. For example, on the \symtestdata{ASV19 test set}, the average EER of systems trained on the \symtrndata{ASV19 training set} (green curve) is the best. The advantage is also observed on the \symtestdata{LA21} and \symtestdata{DF21} test sets, wherein the data or part of the data is from the same domain as the \symtrndata{ASV19 training set} (i.e., the VCTK corpus). Similarly, on the \symtestdata{CodecFake} test set, the gap between the matched and unmatched case is significant. One notable exception is that \symtrndata{ADD23} training set, which does not even perform well on the corresponding test set. 
These results underscore the importance of careful training set selection and motivates the use of multi-condition or pooled training strategies in future work.

\begin{figure*}[htbp]
    \centering
    \includegraphics[width=1\linewidth]{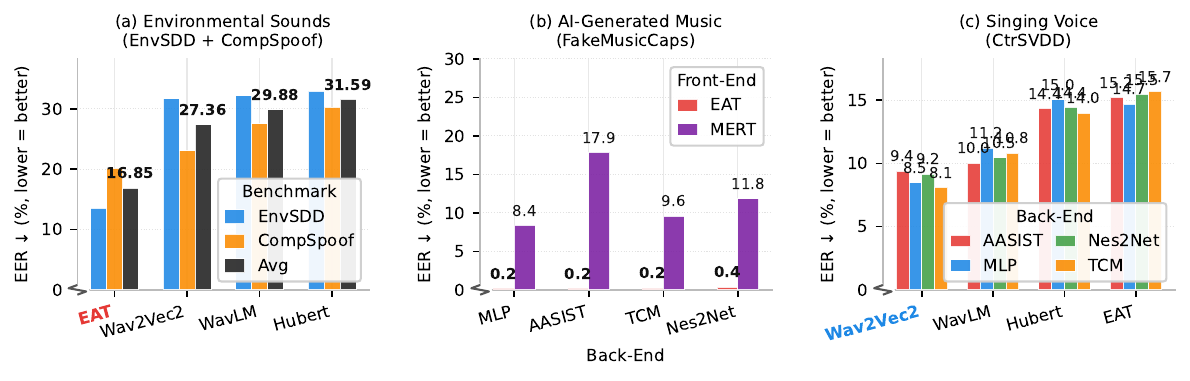}
    \vspace{-7mm}
    \caption{EER on non-speech deepfake detection.}
    \label{fig:beyond_speech}
\end{figure*}

\subsection{Best and worst performing systems}

The best overall system is \textbf{Wav2Vec2 + MLP + CodecFake} with a macro-average EER of $17.16\%$, closely followed by \textbf{Wav2Vec2 + AASIST + CodecFake} ($17.20\%$) and \textbf{Wav2Vec2 + Nes2Net + CodecFake} ($18.20\%$).
The top-10 systems are dominated by the \symfrontend{Wav2Vec2} front-end with \symtrndata{CodecFake} or \symtrndata{HABLA} training, reinforcing the complementary importance of front-end and training set selection.
The worst-performing system is \textbf{HuBERT + MLP + ADD23} ($\text{EER} = 60.4\%$), followed by \textbf{WavLM + MLP + ADD23} ($60.1\%$), the \symtrndata{ADD23} training condition is solely responsible for the worst performers regardless of the front-end or back-end chosen.

\subsection{Beyond speech: song, music, and environmental deepfake sound detection}

\deepfense{} natively supports \textbf{environmental sound}, \textbf{music}, and \textbf{singing voice} deepfake detection with no pipeline changes.
We evaluate on \symtestdata{EnvSDD} and \symtestdata{CompSpoof} \cite{zhang2025compspoof} (environmental sounds), \symtestdata{FakeMusicCaps} (AI-generated music) \cite{comanducci2025fakemusiccaps}, and \symtestdata{CtrSVDD} (singing voice) \cite{zang2024ctrsvdd}, as shown in Figure.~\ref{fig:beyond_speech}.

Across all tasks, the front-end pre-training domain proves decisive.
For environmental/music sounds, \symfrontend{EAT} dominates (avg.\ EER $16.9\%$), which is a direct reversal of the speech ranking where \symfrontend{Wav2Vec2} led, reflecting \symfrontend{EAT}'s general-audio pre-training.
On FakeMusicCaps, \symfrontend{EAT} achieves near-perfect detection (EER $\approx$\,$0.2\%$), while \symfrontend{MERT}, despite being music-pretrained, reaches only $11.9\%$.
For singing voice (CtrSVDD), \symfrontend{Wav2Vec2} recovers its lead (EER $8.8\%$), with EAT performing worst ($15.3\%$).
Back-end choice remains minor across all tasks (spread ${<}2\%$ EER), consistent with our findings for speech.
These results demonstrate that \deepfense{} expands naturally beyond speech, and that \emph{front-end domain alignment} is the decisive design factor for non-speech deepfake detection.

\section{Fairness Study}

Standard evaluation metrics like EER show overall performance, but they may hide biases that affect certain user groups or conditions. For deepfake detection systems to be used in real-world applications, ensuring equitable protection is paramount. In this section, we investigate the fairness of our trained models across three vital dimensions: audio quality, speaker gender, and language. By utilizing the Gini Aggregation
Rate for Biometric Equitability (GARBE) metric~\cite{chouchane2024comparison} integrated in \deepfense{}, we quantify how choices of the model architecture and training data affect model fairness. A GARBE score of $0$ indicates perfect fairness.

To ensure a comprehensive assessment of fairness, we evaluate selected systems on a pooled test set composed of ASV19, LA21, DF21, ASV5, MLAAD, and CodecFake. This aggregation spans legacy and recent TTS and VC attacks, covers both English-only and multilingual conditions, and incorporates synthesis paradigms ranging from earlier parametric systems to modern neural codec-based generators. To isolate the influence of training data from architectural design, we analyze the strongest-performing systems trained on two contrasting corpora, ASV19 and CodecFake. 

Gender fairness analysis leverages the speaker metadata provided with each dataset. For speech quality fairness, perceptual quality scores are computed automatically for every utterance, allowing stratification into objective quality bands without reliance on manual annotations.

\subsection{Speech quality fairness}
\label{sec:fairness_garbe}

\begin{figure*}[htbp]
    \centering
    \includegraphics[width=1\linewidth]{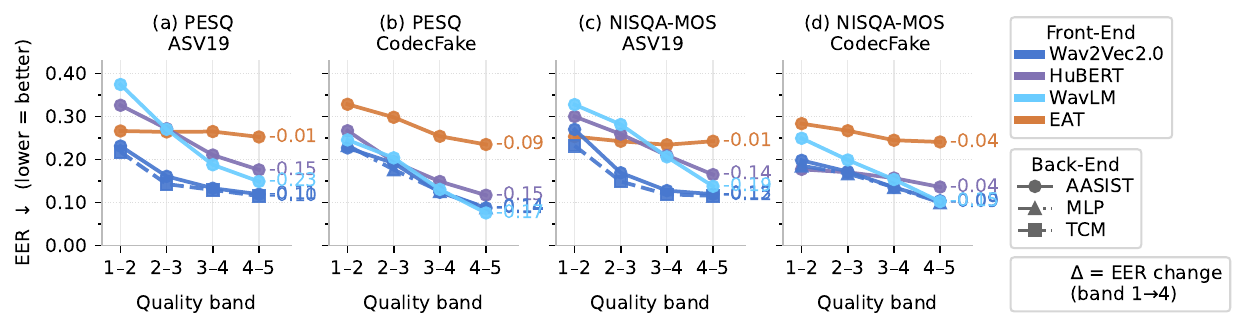}
    \vspace{-7mm}
    \caption{EER Sensitivity to Audio Quality Across Systems trained on ASV19, and CodecFake}
    \label{fig:fairness_quality}
\end{figure*}

To quantify the \emph{fairness} of each detection system with respect to audio quality,
we use two perceptual quality metrics: estimated Perceptual evaluation of speech quality (PESQ) \cite{kumarTorchaudioSquim2023} and NISQA-MOS \cite{mittag21_interspeech}.
Figure~\ref{fig:fairness_quality} visualises EER across quality bands for all system configurations, while Table~\ref{tab:garbe} summarises the resulting GARBE scores. Several observations emerge.



\begin{table}[!t]
\centering
\setlength{\tabcolsep}{3pt}
\caption{GARBE fairness score of each system, computed over audio quality (PESQ and NISQA-MOS), gender, and languages.}
\label{tab:garbe}
\vspace{-2mm}
\resizebox{\columnwidth}{!}{
\begin{tabular}{ll l cccc}
\toprule
         &           &          & \multicolumn{4}{c}{GARBE} \\ \cmidrule{4-7}
Training & Front-End & Back-End & PESQ & NISQA-MOS & Gender & Language\\
\midrule
ASV19 & EAT & AASIST & \textbf{0.0135} & \textbf{0.0196} & 0.1460 & \textbf{0.1886} \\
 & Hubert & AASIST & 0.1735 & 0.1626 & \textbf{0.1160} & 0.2397 \\
 & Wav2Vec2 & AASIST & 0.1880 & 0.2396 & 0.1940 & 0.2147 \\
 & Wav2Vec2 & TCM & 0.1769 & 0.2056 & 0.2110 & 0.2630 \\
 & WavLM & AASIST & 0.2579 & 0.2258 & 0.1450 & 0.2841 \\
\midrule
CodecFake & EAT & AASIST & \textbf{0.0970} & \textbf{0.0482} & 0.3150 & \textbf{0.1497} \\
 & Hubert & AASIST & 0.2271 & 0.0712 & \textbf{0.2940} & 0.1735 \\
 & Wav2Vec2 & AASIST & 0.2576 & 0.1745 & 0.3070 & 0.2032 \\
 & Wav2Vec2 & MLP & 0.2594 & 0.1633 & 0.3300 & 0.2214 \\
 & WavLM & AASIST & 0.2965 & 0.2309 & 0.3360 & 0.1795 \\
\bottomrule
\end{tabular}}
\end{table}

First, systems using the \symfrontend{EAT} front-end consistently achieve the lowest GARBE values across both quality metrics and both training sets
(e.g.\ $0.0135$ on PESQ/ASV19, $0.0482$ on NISQA-MOS/CodecFake),
indicating that \symfrontend{EAT}-based systems are substantially more equitable with respect to audio quality than any other front-end tested. This is notable because \symfrontend{EAT} does not achieve the lowest absolute EER in standard
evaluations, yet its performance remains remarkably stable across quality bands, suggesting that it captures speech characteristics that are less sensitive to recording conditions.
Despite delivering competitive detection performance \symfrontend{Wav2Vec2} systems exhibit substantially higher GARBE scores, revealing a fairness, performance trade-off.

Then, the choice of back-end classifier has a comparatively minor effect on GARBE relative to the front-end, consistent with the view that the feature extractor is the primary driver of quality-band fairness rather than the classification head.

Finally, systems trained on \symtrndata{CodecFake} generally exhibit higher GARBE scores than those trained on \symtrndata{ASV19}. This can be understood by considering the nature of each corpus: \symtrndata{ASV19} was constructed using TTS systems available at the time of its creation,
which produced speech of comparatively lower and more uniform perceptual quality; the resulting training distribution does not sensitise models to quality variation. \symtrndata{CodecFake}, by contrast, is built on recent neural codec tokenizers (\emph{e.g.}\ EnCodec~\cite{de2023encodec}, SoundStream~\cite{zeghidourSoundStream2022}) that produce high-fidelity, perceptually realistic speech. 
Models trained on \symtrndata{CodecFake}, therefore, learn features tied to high-quality synthesis artefacts, which may generalise poorly to lower-quality recordings and thereby amplify performance disparities across quality bands.

Taken together, these findings reveal a \emph{fairness--performance trade-off} that is driven by front-end design: \symfrontend{EAT} stands out as particularly fairer in quality-diverse environments, while \symfrontend{WavLM} and \symfrontend{Wav2Vec2}, despite their strong average EER, should be used with caution in settings where audio quality is heterogeneous.

\subsection{Gender fairness analysis}
\label{sec:fairness_gender}

We evaluate whether the detection systems exhibit disparate performance across gender groups by measuring EER separately for female and male speakers. We report the gap $\Delta = \text{EER}_\text{F} - \text{EER}_\text{M}$, where a positive value indicates that female speakers are disadvantaged, and a negative value that male speakers are.
Table~\ref{tab:gender_eer} shows per-gender EER across all configurations, and the Gender column of Table~\ref{tab:garbe} summarises the GARBE scores.

\begin{table}[!t]
\centering
\caption{EER by gender for each system. F = Female, M = Male, $\Delta = \text{EER}_\text{F} - \text{EER}_\text{M}$.}
\label{tab:gender_eer}
\vspace{-2mm}
\setlength{\tabcolsep}{5pt}
\renewcommand{\arraystretch}{1.15}
\resizebox{\columnwidth}{!}{
\begin{tabular}{lll ccc}
\toprule
& & & \multicolumn{3}{c}{EER} \\
\cmidrule(lr){4-6}
Training & Front-End & Back-End & F & M & $\Delta$ \\
\midrule
ASV19     & EAT      & AASIST & 0.584 & 0.578 & $+0.005$ \\
          & Hubert   & AASIST & 0.447 & 0.321 & $+0.126$ \\
          & Wav2Vec2 & AASIST & 0.257 & 0.183 & $+0.075$ \\
          & Wav2Vec2 & TCM    & 0.275 & 0.200 & $+0.074$ \\
          & WavLM    & AASIST & 0.455 & 0.347 & $+0.108$ \\
\midrule
CodecFake & EAT      & AASIST & 0.135 & 0.182 & $-0.047$ \\
          & Hubert   & AASIST & 0.134 & 0.155 & $-0.021$ \\
          & Wav2Vec2 & AASIST & 0.142 & 0.162 & $-0.020$ \\
          & Wav2Vec2 & MLP    & 0.164 & 0.168 & $-0.004$ \\
          & WavLM    & AASIST & 0.171 & 0.174 & $-0.003$ \\
\bottomrule
\end{tabular}}
\end{table}

\begin{figure*}[!t]
  \centering
  \includegraphics[width=1\linewidth]{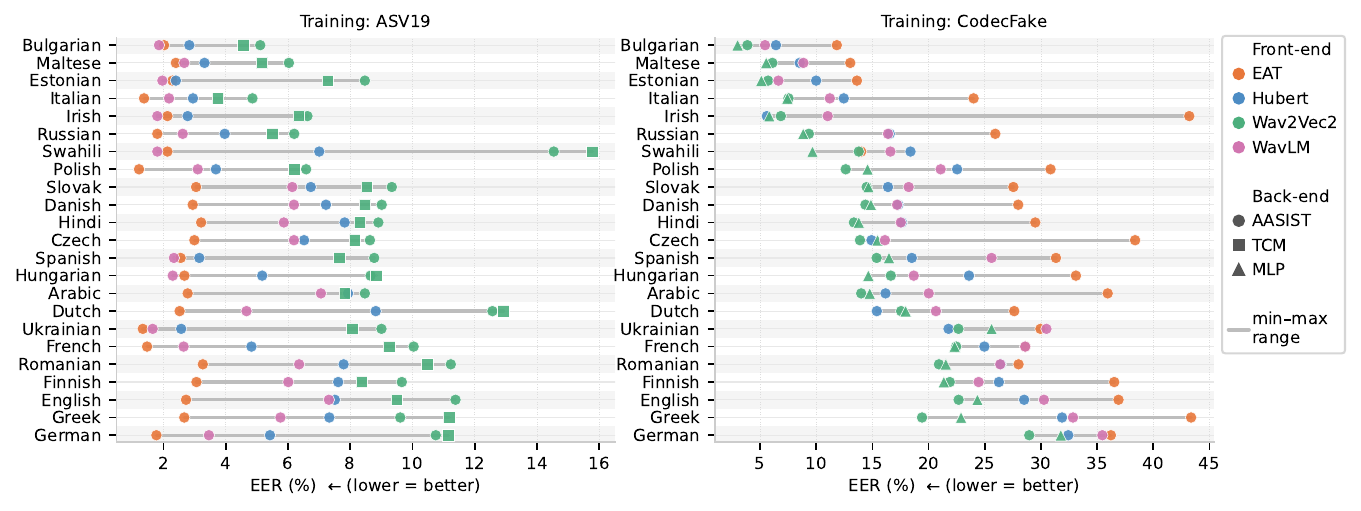}
  \vspace{-5mm}
  \caption{EER~(\%) per language for all system configurations trained on
           ASVspoof~2019 (left) and CodecFake (right).
           Languages are sorted by mean EER (lowest at top).
           Each dot is one system; colour encodes front-end; marker shape
           encodes back-end.
           Grey bars show the min--max EER range across all configurations.}
  \label{fig:lang_dotplot}
\vspace{-0.5cm}
\end{figure*}

All five \symtrndata{ASV19}-trained configurations yield $\Delta > 0$, confirming that female speakers consistently suffer higher EER than male speakers.
The bias is most pronounced for \symfrontend{Hubert} ($\Delta = +0.126$) and \symfrontend{WavLM}
($\Delta = +0.108$); even the best-performing system (\symfrontend{Wav2Vec2}, $\Delta = +0.075$) shows a non-trivial gap. 


A striking reversal occurs when training on \symtrndata{CodecFake}: all five configurations yield $\Delta < 0$, meaning male speakers bear the higher EER. The gap is most severe for \symfrontend{EAT} ($\Delta = -0.047$), while \symfrontend{Wav2Vec2}+\symbackend{MLP} and \symfrontend{WavLM} show near-parity ($\Delta \approx -0.003$). This inversion is consistent with the extreme gender imbalance in \symtrndata{CodecFake}: female utterances account for approximately 80.9\% of the corpus, leaving male speakers severely under-represented during training. Consequently, the model's decision boundary is calibrated toward the majority class, systematically over-rejecting genuine male speech.

The rightmost column of Table~\ref{tab:garbe} shows that \symtrndata{ASV19}-trained systems achieve lower GARBE scores overall (range $0.116$--$0.211$) than \symtrndata{CodecFake}-trained systems (range $0.294$--$0.336$), despite the latter having smaller absolute EER gaps. 


The consistent reversal of bias direction between corpora strongly implicates \emph{training data gender imbalance}, not model architecture, as the primary driver of gender unfairness.

\subsection{Language fairness analysis}
\label{sec:fairness_language}

We evaluate cross-lingual fairness across \textbf{23 languages} from MLAAD,
spanning Germanic, Romance, Slavic, Celtic, Finno-Ugric, Hellenic, Semitic,
Indo-Aryan, and Bantu families (Figure~\ref{fig:lang_dotplot}, Table~\ref{tab:garbe}). 

Under \symtrndata{ASV19} training, \symfrontend{EAT} achieves by far the narrowest per-language spread ($2.0\%$ range), while \symfrontend{Wav2Vec2} variants are the most variable ($9.7$--$12.0\%$), confirming \symfrontend{EAT} as the most equitable front-end, though it has consistently low performance on pooled speech datasets. A notable exception is \textbf{Swahili}, which is problematic only for \symfrontend{Wav2Vec2} ($14.5$--$15.8\%$) while remaining trivial for \symfrontend{EAT} and \symfrontend{WavLM} ($\approx 2\%$).

\symtrndata{CodecFake} training substantially amplifies cross-lingual disparities across all front-ends. Unlike the quality and gender axes, no single front-end maintains a narrow language spread here: per-language EER ranges reach $25$--$31\%$ across all configurations. German is the hardest language for all systems ($29$--$36\%$), while Bulgarian, Estonian, and Maltese remain consistently easy regardless of front-end, suggesting their spoof voices retain detectable artifacts. Irish presents the starkest within-language cross-system contrast: trivial for \symfrontend{Hubert} and \symfrontend{Wav2Vec2} ($\approx 6\%$) but among the hardest for \symfrontend{EAT} ($43\%$). Overall, language-group fairness under \symtrndata{CodecFake} is more uniformly poor than under \symtrndata{ASV19}, pointing that not only the front-end but corpus-level diversity are the key missing ingredients for cross-lingual robustness.

\section{Discussion}
Here we summarize the findings from the  experimental evaluation with around 100 distinct systems built using the \deepfense{} toolkit.
First, the choice of front-end turned out to be critical. Across the 13 evaluation datasets, \symfrontend{Wav2Vec2} performed the best, achieving a macro-average EER of 25.5\%.  However, the front-end superiority is domain-dependent. While \symfrontend{Wav2Vec2} leads in speech and singing voice detection, \symfrontend{EAT} performed better in environmental sound deepfake detection due to its specialized general-audio pre-training. 
In contrast, the back-end classifier has a negligible impact on overall performance. The EER spread between the best back-end (\symbackend{AASIST} at 30.3\%) and the simplest (\symbackend{MLP} at 31.1\%) was less than 1\%. 

Second, the training dataset choice proved to be influential as well. \symtrndata{CodecFake} was identified as the most transferable training set, achieving the lowest macro-average EER (22.3\%). This suggests that the artifacts introduced by the latest DNN-based codecs provide useful cues that help models identify artifacts across various generation methods. Conversely, models trained on the \symtrndata{ADD23 dataset} suffered from generalization failure (EER 50.8\%). 

Finally, our comprehensive fairness analysis underscores a critical vulnerability in current detection paradigms. While models like \symfrontend{Wav2Vec2} achieve top-tier average performance, they exhibit significant biases across audio quality bands, speaker genders, and languages, often driven by demographic and qualitative imbalances in the training data (e.g., the heavy female skew in CodecFake). In contrast, models with general-audio pre-training, such as \symfrontend{EAT}, demonstrate remarkable equity. This highlights an urgent need for the community to prioritize fairness: a system cannot be deemed truly robust if its protection fails disproportionately for specific linguistic or demographic groups.



\section{Conclusion}
\label{sec:conclusion}

In this work, we presented \deepfense{}, a PyTorch-based toolkit for speech deepfake detection with a configuration-driven design that lowers the barrier to reproducible experimentation. Across a large-scale evaluation spanning 13 datasets and three fairness axes, two factors consistently dominate both performance and group-level equity: front-end selection and training data. \symfrontend{Wav2Vec2} leads on average EER for speech domains, yet exhibits the widest disparities across audio quality, gender, and language; \symfrontend{EAT}, by contrast, sacrifices some average performance in exchange for substantially more equitable behaviour across all groups. Training data tells a parallel story: \symtrndata{CodecFake} transfers best across generation methods, while corpora with demographic or qualitative imbalances directly propagate unfairness into the deployed model.

Current limitations include the absence of a multi-dataset training pipeline across corpora, which would be necessary to combine the complementary knowledge of existing datasets without inheriting their individual biases. Detection also remains the only supported task; and extending \deepfense{} to partial deepfake localization and source tracing are natural next steps. Beyond these technical directions, we plan to grow a community around \deepfense{} where researchers can contribute new architectures, loss functions, and training recipes, and to align with production-oriented ecosystems to ensure that reproducible research findings translate into deployable systems.



\section{References}
\printbibliography

\end{document}